\documentclass[prd,showpacs,preprintnumbers,amsmath,amssymb]{revtex4}

\usepackage{graphicx}
\usepackage{dcolumn}
\usepackage{bm}

\def\be{\begin{eqnarray}}
\def\ee{\end{eqnarray}}
\def\bea{\begin{eqnarray}}
\def\eea{\end{eqnarray}}

\def\0T{{\bf 0}_\perp}
\begin{document}


\title{Inclusive Single-Spin Asymmetries, Quark-Photon, and Quark-Quark
Correlations}

\author{Matthias Burkardt}
 \affiliation{Department of Physics, New Mexico State University,
Las Cruces, NM 88003-0001, U.S.A.}
\date{\today}

\begin{abstract}
We consider quark-photon correlations that have been proposed as a source for single-spin asymmetries in inclusive
deep-inelastic scattering. A new sum rule for these correlators is derived and its phenomenological consequences are discussed. The results are interpreted within the context of an 
intuitive 'electrodynamic lensing' picture. 
\end{abstract}

\maketitle
\section{Introduction}
Recent inclusive Deep Inelastic Scattering (DIS) experiments on a transversely polarized target in Hall A at Jefferson Lab showed for the first time a (small) Single-Spin Asymmetry (SSA) for the scattered electron
\cite{Katich}. 
As such an asymmetry has to vanish in single photon exchange, these measurements potentially reveal important
information about quark correlations in the nucleon.

Since the leading order (in $\alpha_{QED}$) SSA arises from the interference between the one photon exchange and the
two photon exchange amplitude, it is unlikely that both
photons involve large momentum transfers with different
quarks as this would lead to a more complex final state (e.g. two jets) that
would not interfere much with a typical one photon exchange event.

Thus two hard photon exchanges would arise dominantly from events where both photons couple to the same quark line. The resulting effective interaction has been estimated
in Ref. \cite{Metz} and will not be considered in this work. The other possibility for a large momentum transfer on the electron are
 processes where one of the exchanged photons is hard and the other one is soft. In this work we will focus on the latter processes, which we not only believe are dominant for inclusive SSAs but also carry
information about the spatial structure of the hadron as we will explore.

\section{Quark Photon Quark Correlator}
In a DIS process, the transverse position of the
scattered electron should be very close to that of the struck quark. One may thus estimate the effect from the initial and final state interactions of the electron by correlating the
leading twist quark density with the electromagnetic field
strength tensor at the same transverse position. 
This observation  motivates to consider \cite{Metz}
\be
-M\epsilon_T^{ij}S_T^j  F_{FT}^q \equiv 
\int \frac{d\xi^-d\zeta^-}{2(2\pi)^2}e^{ixP^+\xi^-}
\langle P,S|\bar{\psi}^q(0)\gamma^+eF_{QED}^{+i}(\zeta^-)\psi^q(\xi^-)|P,S\rangle,
\label{eq:FFT}
\ee
where $e>0$ is the electric charge. $x$ represents the quark
momentum, which is not changed in this
'soft photon pole' matrix element. If the electromagnetic field strength tensor $F_{QED}^{+i}$ is replaced by its QCD counterpart then (\ref{eq:FFT}) represents the soft gluon pole
matrix element \cite{QS} for the single-spin asymmetry
in Semi-Inclusive DIS (SIDIS). We will make use of this analogy several times.

First we note that although Eq. (\ref{eq:FFT}) represents (up to a minus sign)
the average transverse momentum acquired by the
electron due to ISI and FSI, it would yield the
average transverse momentum of the active quark due to
electromagnetic FSI if we were to multiply by
$\frac{1}{2} e_q$, where $e_u=\frac{2}{3}$ and
$e_d=-\frac{1}{3}$. 
The factor $\frac{1}{2}$ arises here since there is only FSI acting on the quark while the electron experiences both ISI and FSI.

In Ref. \cite{mb:quark} the electromagnetic field appearing
in Eq. (\ref{eq:FFT}) was related to the $\perp$ charge density
as
\be
\int dx^- F^{+i}_{QED}(x^-,{\bf x}_\perp)
= - \int \frac{d^2 {\bf y}_\perp}{2\pi}
\frac{x^i-y^i}{|{\bf x}_\perp -{\bf y}_\perp|^2}\rho({\bf y}_\perp),
\ee
where
\be
\rho({\bf y}_\perp) = e\sum_{q^\prime} e_{q^\prime} \int dy^- \bar{\psi}^{q^\prime}
(y^-,{\bf y}_\perp)\gamma^+\psi^{q^\prime}
(y^-,{\bf y}_\perp)
\ee
is the charge density (integrated over $y^-$).
The average $\perp$ momentum for flavor $q$ 
\be F_{FT}^q\equiv \int dx F_{FT}^q(x,x)\ee
can thus be
expressed as
\be
-M\epsilon_T^{ij}S_T^j  F_{FT}^q \equiv 
\int \frac{d\xi^-d\zeta^-}{4\pi P^+}\int 
\frac{d^2{\bf y}_\perp}{2\pi}\frac{y^i}{{\bf y}_\perp^2}
\langle P,S|\bar{\psi}^q(0)\psi^q(0)
\rho({\bf y}_\perp)|P,S\rangle
\label{eq:FFT2} .
\ee
Note that $q^\prime=q$ does not contribute in (\ref{eq:FFT2}) after integration over $d^2{\bf y}_\perp$, i.e. when for example a $u$ quark is struck, the average transverse momentum due to electromagnetic
ISI/FSI is only from fields caused by $d$ (or $s$ and heavier) quarks and {\it vice versa}.
Furthermore, the average $\perp$ momentum of $u$ quarks due to electromagnetic
FSI with $d$ quarks is equal and opposite to 
the average $\perp$ momentum of $d$ quarks due to electromagnetic
FSI with $u$ quarks. 
As a corollary, one finds 
the 'sum rule'  \cite{mb:quark}
\be
\frac{2}{3}F_{FT}^u-\frac{1}{3}F_{FT}^d+...=0
\label{eq:sumrule}
\ee
regardless whether the target is a proton or a neutron.
Thus similar to the case for QCD \cite{mb:glue}, the average transverse momentum due to the FSI also vanishes in the
abelian case, provided one sums over all charged constituents. Note that if one neglects strange or heavier quarks, then the sum rule implies that 
$F_{FT}^u$ and $F_{FT}^d$ must have the same
sign so that they can sum to zero after weighting with $e_q$
\be
F_{FT}^d=2F_{FT}^u.
\ee
This result should apply to any target, i.e. the transverse 
momentum asymmetry on the electron should receive
contributions from electromagnetic interactions with $u$ and $d$ quarks that are of the same sign for any given target.
The model asumptions  made in Ref. \cite{Metz},  lead to $F_{FT}^{u/p}$ and $F_{FT}^{d/p}$
having opposite signs, so that the sum rule cannot be satisfied. For the neutron, $F_{FT}^{u/p}$ and $ F_{FT}^{d/p}$ have the same sign, consistent with
(\ref{eq:sumrule}). However, with the choice of coefficients
made in Ref. \cite{Metz} the sum rule does not seem to
be satisfied for the neutron either.

\section{A Model for Quark Photon Correlators}

In Ref. \cite{Metz} it was proposed to estimate
$F_{FT}^q(x,x)$ by taking phenomenological fits \cite{fits} of its 
QCD counterpart $T_F^q(x,x)$ and rescale those as
\be
F_{FT}^q(x,x) =  -e_{q_s}\frac{\alpha_{em}}{2\pi C_F \alpha_s M}gT_F^q(x,x),
\label{eq:rescale}
\ee
where $e_{q_s}$ is the charge of the spectators: for example, if the active quark is a $u$ quark in a proton then
$e_{q_s}=\frac{1}{3}$ in the model of Ref. \cite{Metz}. This results in
quark-photon correlators
\be
F^{u/p}_{FT}(x,x)=-\frac{\alpha_{em}}{6\pi C_F \alpha_s M}gT_F^{u/p}(x,x) \quad\quad\quad
F^{d/p}_{FT}(x,x)=-\frac{2\alpha_{em}}{3\pi C_F \alpha_s M}gT_F^{d/p}(x,x)
\label{eq:metz1}\\
F^{u/n}_{FT}(x,x)=\frac{\alpha_{em}}{3\pi C_F \alpha_s M}gT_F^{d/p}(x,x) \quad\quad\quad
F^{d/n}_{FT}(x,x)=-\frac{\alpha_{em}}{6\pi C_F \alpha_s M}gT_F^{u,p}(x,x).
\label{eq:metzFFT}
\ee
Here charge symmetry has been used to relate neutron matrix elements to those in
the proton
It is evident that (\ref{eq:metzFFT}) violates the above 'sum-rule' (\ref{eq:sumrule}).

The root of this problem is the assumption that  both spectator quarks contribute
equally to the gauge fields in the matrix elements for $\int dx F_{FT}(x,x)$ and $\int dx T_F(x,x)$.
However, that is not the case as
FSI with quarks from the same flavor as the active quark do not contribute to
either one of them \cite{mb:quark,mb:glue}.

$x$-averaged contributions only arise
from  quarks with flavor other than the active quark due to symmetry.
To account for that in the model from Ref. \cite{Metz} one should thus replace (\ref{eq:rescale}) by
\be
F_{FT}^q(x,x) =  -e_{\tilde{q}}\frac{\alpha_{em}}{\pi C_F \alpha_s M}gT_F^q(x,x),
\label{eq:rescale2}
\ee
for the majority flavor ($u$ in $p$ and $d$ in $n$)
which is almost identical to (\ref{eq:rescale}), except that
$e_{\tilde{q}}$ would be the charge of the spectator flavor in the proton/neutron.
For example, in the above example where the active quark
is a $u$ quark in a proton $e_{\tilde{q}}=-\frac{1}{3}$, or
for a $d$ quark in a proton $e_{\tilde{q}}=\frac{4}{3}$
This would result in modified inclusive quark-photon correlators as
\be
F^{u/p}_{FT}(x,x)=\frac{\alpha_{em}}{3\pi C_F \alpha_s M}gT_F^{u/p}(x,x) \quad\quad\quad
F^{d/p}_{FT}(x,x)=-\frac{2\alpha_{em}}{3\pi C_F \alpha_s M}gT_F^{d/p}(x,x)
\label{eq:newFFT1}
\\
F^{u/n}_{FT}(x,x)=\frac{\alpha_{em}}{3\pi C_F \alpha_s M}gT_F^{d/p}(x,x) \quad\quad\quad
F^{d/n}_{FT}(x,x)=-\frac{2\alpha_{em}}{3\pi C_F \alpha_s M}gT_F^{u,p}(x,x).
\label{eq:newFFT2}
\ee
This result agrees with Ref. \cite{Metz} in the case of
minority flavor ($F^{d/p}_{FT}$ \& $F^{u/n}_{FT}$) correlators,
 but differs for the majority flavor, where Ref. \cite{Metz}
multiplies by the net charge of all spectators, but we only
multiply by the net charge of the spectator flavor. Furthermore, we divide for the majority flavor by $\frac{1}{2}C_F\alpha_s$ rather than $C_F\alpha_s$
The latter step is to account for the fact that the QCD FSI interaction between one of the $u$
quarks in the proton with the $d$ quark has a color factor that is only $\frac{1}{2}$
the color factor for the $d$ quark in a proton to interact with the $u$-diquark
pair.

It is easy to verify that these $F_{FT}$
satisfy the above 'sum rule'
(\ref{eq:sumrule}), i.e.
\be
\frac{2}{3}\int dx F^{u/N}(x,x) -\frac{1}{3}\int dx
F^{d/N}(x,x)=0,
\ee
for both $N=n,p$,
 provided the $T^{q/p}_F$ saturate the corresponding sum rule
$\int dx T_F^{u,p}(x,x) + \int dx T_F^{d,p}(x,x)=0$.

\section{Discussion}
In the model proposed in Ref. \cite{Metz} the inclusive SSA is proportional to 
\be
\sigma_{UT}\propto 4 F_{FT}^{u}+F_{FT}^d.
\label{eq:UT}
\ee
Using the correlators from Eq. (\ref{eq:metz1}) one thus finds
\be
\sigma_{UT}^p\propto -\frac{2\alpha}{3\pi C_F \alpha_sM}g\left(T_F^{u/p}+
T_F^{d/p}\right).
\ee
Recent extractions of the Sivers function from SIDIS data indicates that
$T_F^{d/p}\approx -T_F^{u/p}$, which is also consistent with a very small
Sivers function on a deuterium iarget result in a very small result for
$\sigma_{UT}^p$ with the correlators from Eq. (\ref{eq:metz1}).

In contradistinction, with the new quark-photon-quark correlators (\ref{eq:newFFT1}) one finds
\be
\sigma_{UT}^p\propto \frac{2\alpha}{3\pi C_F \alpha_sM}g\left(2T_F^{u/p}-
T_F^{d/p}\right),
\ee
where no cancellation occurs provided $T_F^{u/p}$ and $T_F^{d/p}$ have opposite signs.

For the neutron, the resulting change of the asymmetry with the new $F_{FT}$ is small, 
since only
$F_{FT}^{d/n}$ has changed (increased by factor 4) compared to Ref. \cite{Metz}.
In the asymmetry, $F_{FT}^{d/n}$ gets
multiplied by the charge squared of the down quark and thus the asymmetry increases by only about $50\%$. With the values from Eq. (\ref{eq:newFFT2}) one finds
\be
\sigma_{UT}^n\propto \frac{2\alpha}{3\pi C_F \alpha_sM}g\left(2T_F^{d/p}-
T_F^{u/p}\right),
\ee
which is equal and opposite to that of the proton provided one makes the additional
assumption that $T_F^{d/p}\approx T_F^{u/p}$. Of course, the
asymmetries would still be numerically larger for the neutron but only because
one divides by the total cross section.
Nevertheless, our results predict a significant cancellation for deuterium as
\be
\sigma_{UT}^d\propto \frac{2\alpha}{3\pi C_F \alpha_sM}g\left(T_F^{d/p}+
T_F^{u/p}\right).
\ee
While the above relation $\sigma_{UT}^p\approx -\sigma_{UT}^n$ was derived on the
basis of what we believe is an improved version of the model from Ref. \cite{Metz},
it can also be derived using only charge symmetry and Eq. (\ref{eq:UT}), i.e. neglecting
$s$ and heavier quarks and assuming that the cross section asymmetry is
described by the quark photon correlator.
The key observation is the fact that (after integrating over $x$)
photons that contribute to $F^u_{FT}$ cannot
originate from $u$ quarks and similarly for $d$ quarks. Combining this result
with the approximate assumption that only $u$ and $d$ quarks play a role in these matrix elements this implies that photons that contribute to $F^u_{FT}$ can only
originate from $d$ quarks and {\sl vice versa}. In combination with charge
symmetry this allows us to relate proton and neutron matrix elements after rescaling by the charge of the quark flavor from which the photons originated as
\be
\frac{1}{-\frac{1}{3}}F_{FT}^{u/p} = \frac{1}{\frac{2}{3}}F_{FT}^{d/n}\\
\frac{1}{\frac{2}{3}}F_{FT}^{d/p} = \frac{1}{-\frac{1}{3}}F_{FT}^{u/n}
\ee
or
\be
-2F_{FT}^{u/p} = F_{FT}^{d/n}\\
F_{FT}^{d/p} = -2F_{FT}^{u/n}
\ee

\begin{figure}
\includegraphics[scale=.75]{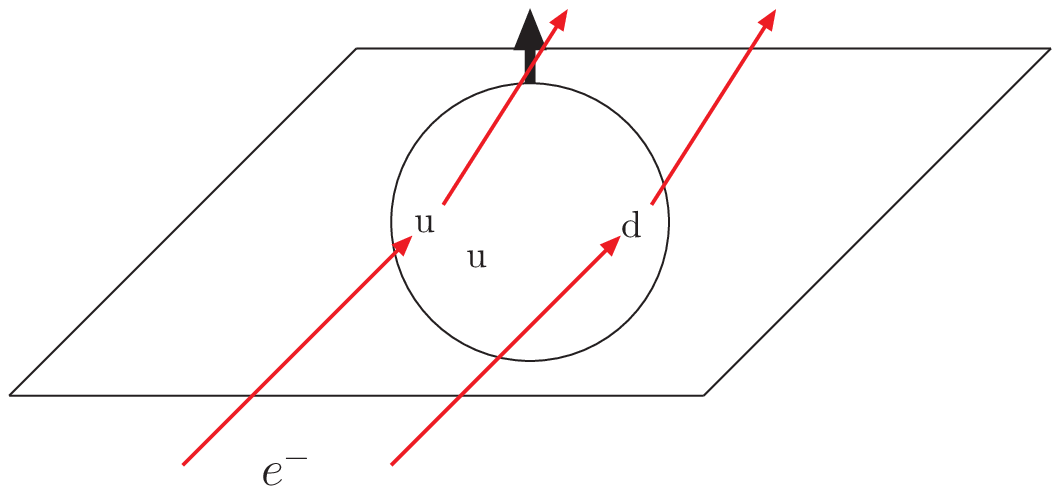}
\includegraphics[scale=.75]{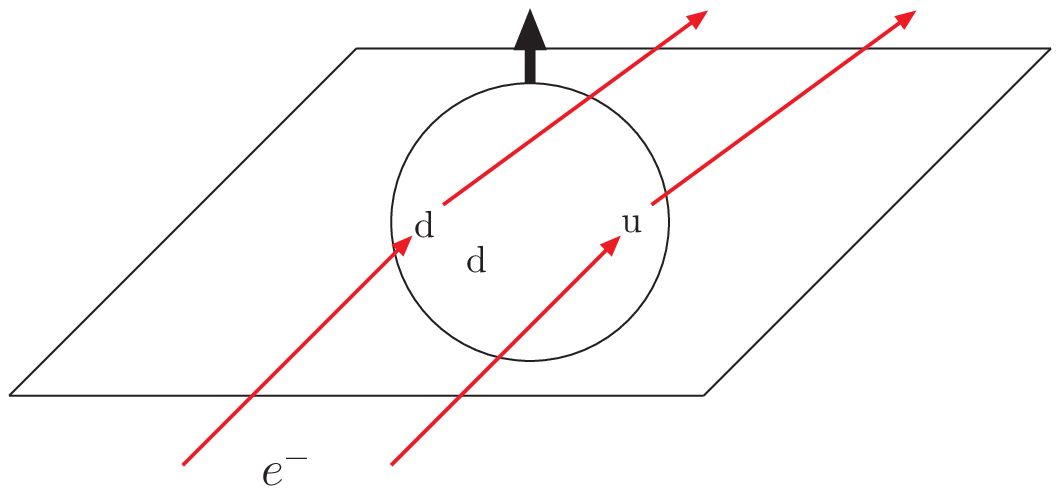}
\caption{Inclusive SSA in the lensing picture. When the $e^-$ knocks out a $u$ quark in a proton  that is polarized up (left picture), this happens preferentially on the left side of the nucleon. The repulsive force from the spectator $d$ quark on average excerts a force on the $e^-$ to the left. When the $e^-$ knocks out the $d$ quark, the attractive force from the $u$ quarks is on average also to the left. Fot a neutron target (right picture) the forces are preferentially to the right.
}
\label{fig:lensing}
\end{figure}

As illustrated in Fig. \ref{fig:lensing} the same sign contribution for scattering from $u$ {\sl vs.} $d$ quarks can be understood
from a mechanism similar to the 'lensing mechanism' proposed in Ref. \cite{lensing}.
For a transversely polarized nucleon the virtual hard photon sees $u$ quarks shifted
towards one side of the nucleon and $d$ quarks to the other. When
the $e^-$ knocks out a $u$ quark, it is repelled by the negatively charged $d$ quarks
on the other side of the nucleon. As explained above the average transverse momentum from interactions with spectator $u$ quarks is zero.
On the other hand, when the $e^-$ knocks out a $d$ quark it is attracted by the
positively charged $u$ quarks. However, since the $u$ and $d$ distributions in a
transversely polarized nucleon are deformed in opposite directions, the
net force from the spectators on the $e^-$ is in both cases (knocking out
$u$ or $d$ quarks) in the same direction, i.e. there should not be a cancellation
between $u$ and $d$ quarks.

For the neutron, the resulting change of the asymmetry is small, since only
$F_{FT}^{d/n}$ has changed (increased by factor 4) compared to Ref. \cite{Metz}.
In the asymmetry, $F_{FT}^{d/n}$ gets
multiplied by the charge squared of the down quark and thus the asymmetry increases by only about $50\%$.

For the proton the change is more significant, as our result for $F_{FT}^{u/p}$ has a sign which differs from that in Ref.\cite{Metz}, and there is
no longer an almost complete cancellation between
$u$ and $d$ contributions to $\sigma_{UT}^p$. Moreover,
since the latter gets multiplied by $e_u^2=\frac{4}{9}$, the resulting
change is quite significant. In fact, we now expect an asymmetry in the proton of the same order of magnitude as that in
the neutron. To see this we consider the
ratio 
\be
\frac{A_{UT}^p}{A_{UT}^n}=
\frac{\sigma_{unp}^n}{\sigma_{unp}^p}
\frac{4 F_{FT}^{u/p}
+ F_{FT}^{d/p}}{4 F_{FT}^{u/n}
+ F_{FT}^{d/n}}
= \frac{\sigma_{unp}^n}{\sigma_{unp}^p}
\frac{2 T_{F}^{u/p}
- T_{F}^{d/p}}{2 T_{F}^{d/p}
- T_{F}^{u/p}}.
\ee
For an order of magnitude estimate, we
approximate $T_F^{d/p}\approx
-T_F^{u/p}$, yielding 
\be
\frac{A_{UT}^p}{A_{UT}^n}\approx -
\frac{\sigma_{unp}^n}{\sigma_{unp}^p},
\ee
i.e. a suppression only due to the fact that the
unpolarized cross section is larger for a proton target.
More detailed estimates for the asymmetries can be found in 
Ref. \cite{Tareq}.

The observation that  $\sigma_{UT}^p\approx \sigma_{UT}^n$
should motivate to revisit $A_{UT}^p$ with
better statistics than obtained in the {\sc Hermes} analysis \cite{hermes} where the measured asymmetry is consistent with zero.

\section{Summary}
We have developed a revised model for the quark-photon-quark correlator
relevant for inclusive transverse single-spin asymmetries for scattering of unpolarized electrons from a transversely polarized target. In the model
single-spin asymmetries from SIDIS experiments are used as an input and
after rescaling by the appropriate coupling constants applied to inclusive SSAs. The novel feature in this work is that if an electron knocks out for example a $u$ quark then the average transverse momentum from interactions of that electrons with $u$ quarks vanishes (and similar for $d$ quarks). As a result the quark-photon-quark correlators now observe sum rules analogous
to similar sum rules for quark-gluon-quark correlators. The immediate consequence of this modification is that $\sigma_{UT}$ for a proton is now predicted to be on the same order as that of the proton, and proton asymmetries are only suppressed by the larger total cross sections.

However, beyond this revised prediction for $A_{UT}^p$, our results have much more general implications for spectator models for ISI/FSI. More specifically, our analysis exhibited an issue for the ISI/FSI applied to the majority flavor. In diquark spectator models the spectators are lumped into a single diquark and the ISI/FSI are estimated typically by considering one-gluon exchange interactions with a diquark rather than the spectator quarks individually. For $u$ quarks interacting with a $ud$ diquark, this
implies that the ISI/FSI from interactions with the spectator $u$ quark is the same as that with the spectator $d$ quark. However, from the symmetry of the interaction, the average transverse momentum acquired from interactions with the spectator $u$ quark should be zero \cite{mb:quark}, which is consistent only if the net effect from interactions with the $ud$ diquark vanish, i.e. no net SSA from FSI/ISI for the majority flavor. However, to satisfy the transverse momentum sum rule the
net effect on the $d$ quarks would then also have to be zero, which illustrates that diquark spectator models for the SSAs have an intrinsic conflict with the transverse momentum sum rule.

{\bf Acknowledgements:}
I would like to thank L.~Gamberg, A.~Metz, M.~Schlegel, and W.~Vogelsang for stimulating 
discussions. This work was supported by the DOE under grant number 
DE-FG03-95ER40965. 

\end{document}